\documentclass{eas}
\usepackage{graphicx}
\usepackage{natbib}
\usepackage{amssymb}
\begin{document}

\title{Lithium abundances in exoplanet host stars as test of planetary formation scenarii}

\runningtitle{Castro \etal: Lithium abundances in planet host stars \dots}
\author{M. Castro}
\address{Laboratoire d'Astrophysique de Toulouse et Tarbes}
%
\author{O. Richard}
\address{Groupe de Recherche en Astronomie et Astrophysique du Languedoc}
\author{S. Vauclair}
\sameaddress{1}
%
%
\begin{abstract}
Following the observations of \cite{Israelian04}, we compare different evolutionary models in order to study the lithium destruction processes and the planetary formation scenarii.
\end{abstract}
\maketitle
\section{Introduction}

The observations of \cite{Israelian04} show evidences that the lithium abundances are different in planet host stars than in stars without planets. Planet host stars also show a relative overmetallicity, compared to stars without planet, which could be explained by two different planetary formation scenarii. We have computed two different types of evolutionary models which modelise the two scenarii. The effect of these two hypotheses on the evolution of the surface lithium abundance is presented.

\section{Metallic peculiarity of planet host stars}\label{sec:metallic}

Planet host stars are in average overmetallic by 0.2 dex compared to stars without detected planets (\cite{Santos01}). This peculiarity could be explained in two different ways. The first scenario assumes that the protostellar cloud was overmetallic and that such an overmetallicity favours planetary formation : in this case the star should be overmetallic from the centre to the surface. The second scenario assumes that the protostellar clouds had the same metallicity, and that the surface metallicity excess is due to the accretion of metal-rich matter during the process of planetary formation.

\begin{figure}
 \centering
 \resizebox{6.5cm}{!}{\includegraphics{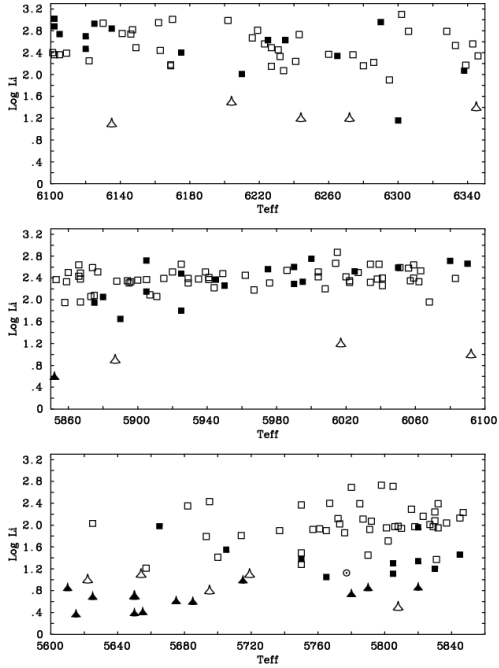}}
  \caption{Lithium versus effective temperature for stars with planets (filled squares) and the comparison sample of Chen et al. (empty squares). Upper limits are filled (planet hosts) and empty (comparison sample) triangles. The position of the Sun is indicated. (\cite{Israelian04})}\label{fig:lithium}
\end{figure}

In \cite{Israelian04}, the authors compare two samples : the first one includes 79 planet host stars, and the second one, due to \cite{Chen01}, includes 183 stars without detected planets. They point out that planet host stars which have an effective temperature between 5600 and 5850 K, show an underabundance of lithium compared to stars without planets (see Figure \ref{fig:lithium}). 

\section{Models and calibration}\label{sec:models}

We compute the evolution of three types of models with the Toulouse-Geneva evolution code: normal models (with a solar metallicity), overmetallic models (three different models, with $[Fe/H]_{0}$ = 0.18, 0.24 and 0.30 from the centre to the surface), and accretion models (three different models with the same metallicity as the overmetallic ones in their outer layers, i.e. accreted mass of metals of respectively $M_{acc}$ = 1.1, 1.7 and 2.5 $M_{J}$), for different masses (0.95, 0.97, 1.00 and 1.02 $M_{\odot}$). All the models include microscopic diffusion, and take into account the rotation-induced turbulence including the effect of $\mu$-gradients as in \cite{Richard96}.

\section{Results and comments}\label{sec:results}

\begin{figure}
 \centering
 \resizebox{6.5cm}{!}{\includegraphics{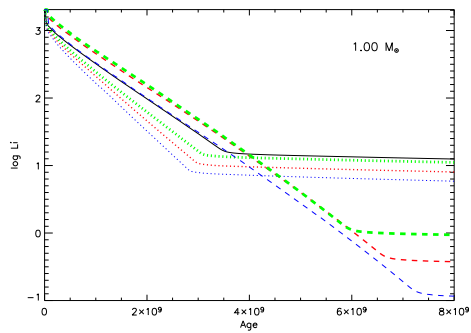}}
  \caption{Lithium evolution for models of 1.00 $M_{\odot}$; solid line : normal metallicity ; dotted lines : overmetallic models ($[Fe/H]_{0}$ = 0.18 (thick green) , 0.24 (red) and 0.30 (thin blue)) ; dashed lines : accretion models ($M_{acc}$ = 1.1 (thick green), 1.7 (red) and 2.5 (thin blue) $M_{J}$).}\label{fig:evolli}
\end{figure}

The evolution of the surface lithium abundances, presented in Figure \ref{fig:evolli} for different models of 1.00 $M_{\odot}$, has two different regimes: a fast decrease due to the rotation-induced turbulence which brings the lithium from the bottom of the convective zone down to the depth where it is burned by nuclear reactions, and a weak decrease when the diffusion-induced helium gradients below the convective zones become large enough to stabilize most of the mixing region. The different slopes during the fast decrease of the surface lithium in the overmetallic models in Figure \ref{fig:evolli} is explained by the temperature profiles: the higher the temperature at a given depth and the stronger the lithium destruction. In the other models, the temperature profiles are almost the same so that the evolution of the surface lithium is similar until the $\mu$-gradient takes over.

\begin{figure}
 \centering
 \resizebox{4.43cm}{5.1cm}{\includegraphics{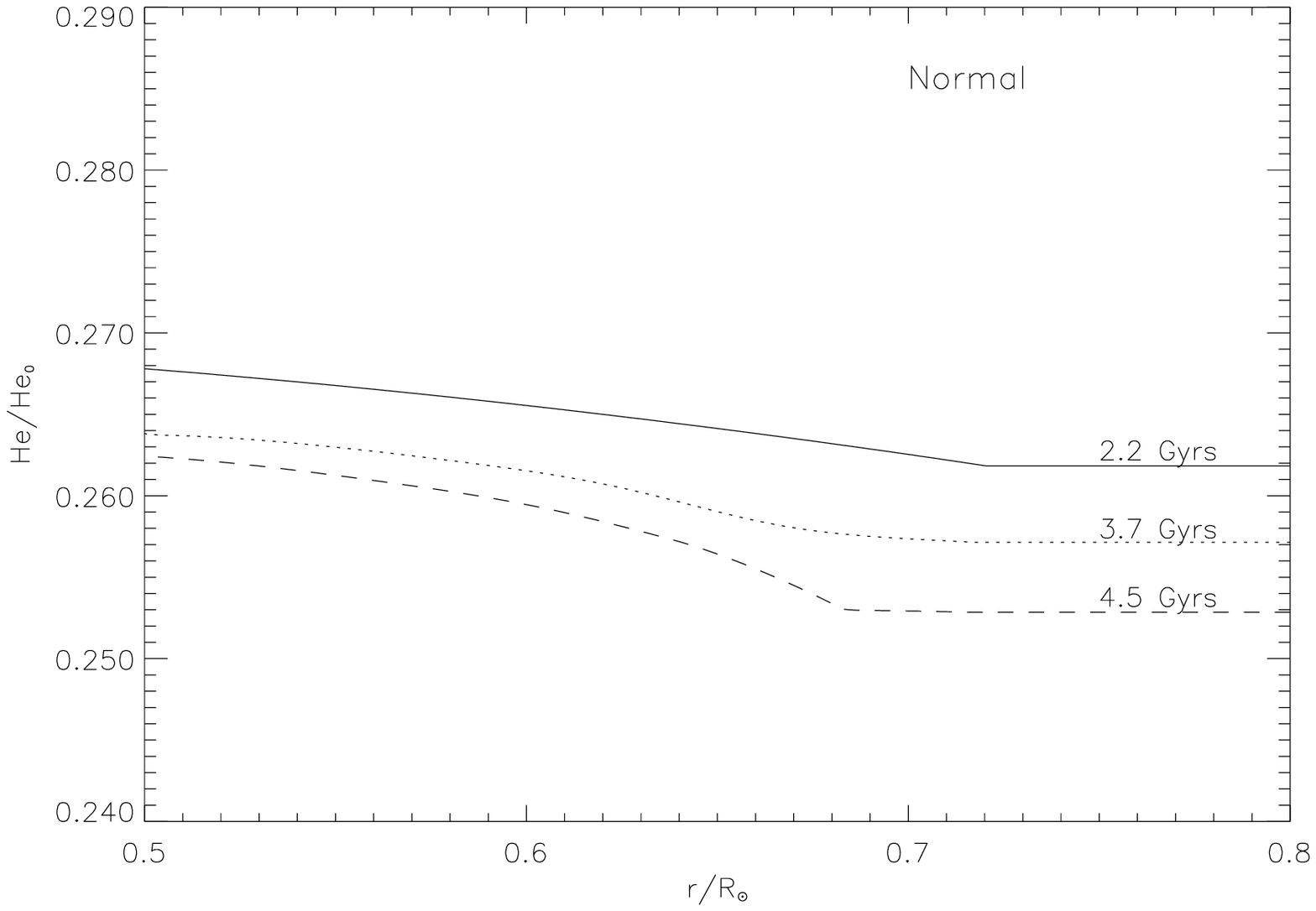}}
 \resizebox{4.43cm}{5.1cm}{\includegraphics{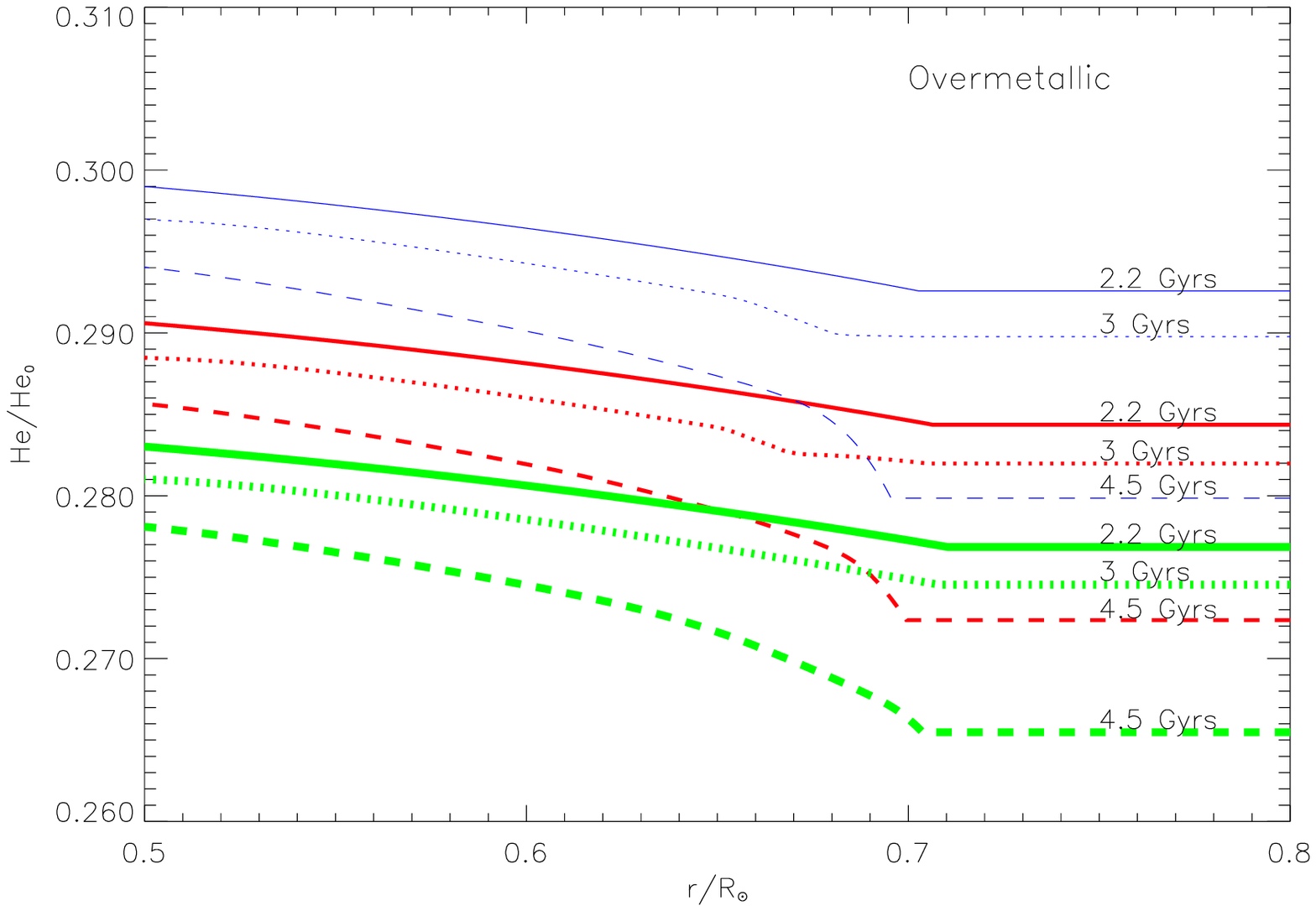}}
 \resizebox{4.43cm}{5.1cm}{\includegraphics{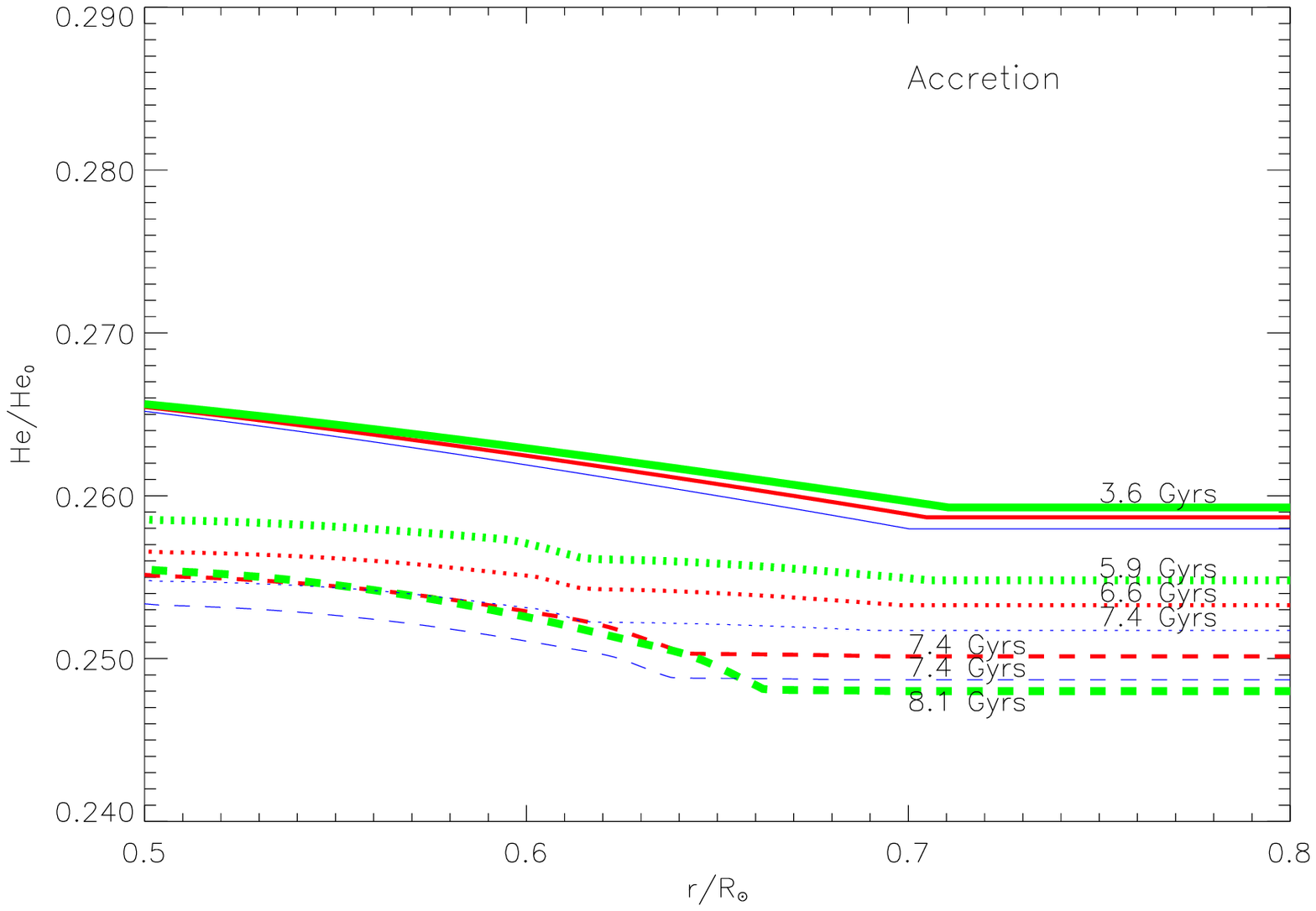}}
  \caption{Helium profiles for the same models as in Figure \ref{fig:evolli} at different ages indicated}\label{fig:profilHe}
\end{figure}

The helium gradients (Figure \ref{fig:profilHe}) are smoother in the accretion models than in the two other cases. Indeed, the creation of an inverse $\mu$-gradient partly contradicts the one due to helium. As a result, turbulence is more efficient and consequently reduces the helium gradient itself : this effect is added to the first one, and, altogether, the mixing process down to the lithium burning layer lasts much longer than in the normal or overmetallic cases.

\section{Conclusion}\label{sec:conclusion}
 
 Following the observations of \cite{Israelian04}, we have modelised several models of solar-type stars in order, on tne one hand to study the lithium destruction mecanisms in these stars, and on the other hand to bring elements of discrimination between the two planetary formation scenarii. We show that overmetallic models have a thicker surface convective zone and so destroy strongly their lithium,but in a short time. The accretion models destroy more slowly lithium, but during a longer time. The continuation of this work is, on the one hand to improve the treatment of the mixing, which is too simplistic, and on the other hand to compute models for hotter stars which do not present the lithium anomaly. 

%
\bibliography{apj-jour, add your bibliography file}
\bibliographystyle{astron}


\end{document}